\documentclass[iop,apjl,useAMS,usenatbib]{emulateapj-rtx4}
\usepackage{graphicx,psfrag}
\usepackage{lscape}
\usepackage{amsmath}
\usepackage{color}
\usepackage{hyperref}
\def\mnras{MNRAS}
\def\aap{A\&A}

\def\aj{AJ}
\def\araa{ARA\&A}
\def\apjl{ApJ}
\def\apj{ApJ}
\def\apjs{ApJS}

\def\nat{Nature}

\keywords{ Galaxies: luminosity function --- Galaxies: star formation --- Infrared: galaxies}

\begin{document}

\shorttitle{Far-IR Emission from DOGs}
\shortauthors{Calanog et al.}

%\title{HerMES: Far-Infrared Emission of Dust Obscured Galaxies in the Bo\"{o}tes Field}

\title{HerMES: The Far-Infrared Emission From Dust Obscured Galaxies}
\author{J.A.~Calanog\altaffilmark{1},
J.~Wardlow\altaffilmark{1},
Hai~Fu\altaffilmark{1,2},
A.~Cooray\altaffilmark{1,3},
R.J.~Assef\altaffilmark{4},
J.~Bock\altaffilmark{3,4},
C.M.~Casey\altaffilmark{5},
A.~Conley\altaffilmark{6},
D.~Farrah\altaffilmark{7,8},
E.~Ibar\altaffilmark{9},
J.~Kartaltepe\altaffilmark{10},
G.~Magdis\altaffilmark{11},
L.~Marchetti\altaffilmark{12,13},
S.J.~Oliver\altaffilmark{7},
I.~P{\'e}rez-Fournon\altaffilmark{14,15},
D.~Riechers\altaffilmark{3},
D.~Rigopoulou\altaffilmark{16,11},
I.G.~Roseboom\altaffilmark{7,17},
B.~Schulz\altaffilmark{3,18},
Douglas~Scott\altaffilmark{19},
M.~Symeonidis\altaffilmark{20},
M.~Vaccari\altaffilmark{13,21},
M.~Viero\altaffilmark{3},
M.~Zemcov\altaffilmark{3,4}}
\altaffiltext{1}{Dept. of Physics \& Astronomy, University of California, Irvine, CA 92697}
\altaffiltext{2}{Department of Physics and Astronomy, University of Iowa, Van Allen Hall, Iowa City, IA 52242}
\altaffiltext{3}{California Institute of Technology, 1200 E. California Blvd., Pasadena, CA 91125}
\altaffiltext{4}{Jet Propulsion Laboratory, 4800 Oak Grove Drive, Pasadena, CA 91109}
\altaffiltext{5}{Institute for Astronomy, University of Hawaii, 2680 Woodlawn Drive, Honolulu, HI 96822}
\altaffiltext{6}{Center for Astrophysics and Space Astronomy 389-UCB, University of Colorado, Boulder, CO 80309}
\altaffiltext{7}{Astronomy Centre, Dept. of Physics \& Astronomy, University of Sussex, Brighton BN1 9QH, UK}
\altaffiltext{8}{Department of Physics, Virginia Tech, Blacksburg, VA 24061}
\altaffiltext{9}{Instituto de Astrof\'isica. Facultad de F\'isica. Pontificia Universidad Cat\'olica de Chile. Casilla 306, Santiago 22, Chile}
\altaffiltext{10}{National Optical Astronomy Observatory, 950 North Cherry Avenue, Tucson, AZ 85719}
\altaffiltext{11}{Department of Astrophysics, Denys Wilkinson Building, University of Oxford, Keble Road, Oxford OX1 3RH, UK}
\altaffiltext{12}{Department of Physical Sciences, The Open University, Milton Keynes MK7 6AA, UK}
\altaffiltext{13}{Dipartimento di Fisica e Astronomia, Universit\`{a} di Padova, vicolo Osservatorio, 3, 35122 Padova, Italy}
\altaffiltext{14}{Instituto de Astrof{\'\i}sica de Canarias (IAC), E-38200 La Laguna, Tenerife, Spain}
\altaffiltext{15}{Departamento de Astrof{\'\i}sica, Universidad de La Laguna (ULL), E-38205 La Laguna, Tenerife, Spain}
\altaffiltext{16}{RAL Space, Rutherford Appleton Laboratory, Chilton, Didcot, Oxfordshire OX11 0QX, UK}
\altaffiltext{17}{Institute for Astronomy, University of Edinburgh, Royal Observatory, Blackford Hill, Edinburgh EH9 3HJ, UK}
\altaffiltext{18}{Infrared Processing and Analysis Center, MS 100-22, California Institute of Technology, JPL, Pasadena, CA 91125}
\altaffiltext{19}{Department of Physics \& Astronomy, University of British Columbia, 6224 Agricultural Road, Vancouver, BC V6T~1Z1, Canada}
\altaffiltext{20}{Mullard Space Science Laboratory, University College London, Holmbury St. Mary, Dorking, Surrey RH5 6NT, UK}
\altaffiltext{21}{Astrophysics Group, Physics Department, University of the Western Cape, Private Bag X17, 7535, Bellville, Cape Town, South Africa}%\pagerange{\pageref{firstpage}--\pageref{lastpage}} \pubyear{2011}

%\maketitle

\label{firstpage}

\begin{abstract}
Dust-obscured galaxies (DOGs) are a UV-faint, infrared-bright galaxy population that reside at $z\sim2$ and are believed to be in a phase of dusty star-forming and AGN activity. We present far-infrared (far-IR) observations of a complete sample of DOGs in the 2 deg$^2$ of the Cosmic Evolution Survey (COSMOS).  The 3077 DOGs have $\langle z \rangle = 1.9$ $\pm$ $0.3$ and are selected from  $24\,\mu$m and $r^{+}$ observations using a color cut of $r^{+}-[24] \ge 7.5$ (AB mag) and $S_{24} \ge 100\,\mu$Jy. Based on the near-IR spectral energy distributions, 47\% are bump DOGs (star-formation dominated) and 10\% are power-law DOGs (AGN-dominated).  We use SPIRE far-IR photometry from the {\it Herschel} Extragalactic Multi-tiered Survey (HerMES)  to calculate the IR luminosity and characteristic dust temperature for the 1572 (51\%) DOGs that are detected at 250 $\mu$m ($\ge 3\sigma$). For the remaining 1505 (49\%) that are undetected, we perform a median stacking analysis to probe fainter luminosities. {\it Herschel}-detected and undetected DOGs have average luminosities of $(2.8 \pm 0.4) \times 10^{12} \text{L}_{\odot}$ and $(0.77\pm0.08) \times 10^{12} \text{L}_{\odot}$, and dust temperatures of $(33\pm7)$ K and $(37\pm5)$ K, respectively. The IR luminosity function for DOGs with $S_{24} \ge 100\,\mu$Jy is calculated, using far-IR observations and stacking. DOGs contribute $10-30\%$ to the total star formation rate density of the Universe at $z=1.5-2.5$, dominated by $250\,\mu$m detected and bump DOGs. For comparison, DOGs contribute 30$\%$ to the star-formation rate density for all $z=1.5-2.5$ galaxies with $S_{24}\ge100\,\mu$Jy. DOGs have a large scatter about the star-formation main sequence and their specific star-formation rates show that the observed phase of star-formation could be responsible for their total observed stellar mass at $z\sim2$.\end{abstract}
% \begin{keywords}
% Keywords
% \end{keywords}

%%%%%%%%%%%%%INTRODUCTION%%%%%%%%%%%%%%
\section{Introduction}
\label{sec:intro}

The far-infrared (far-IR) luminosities of Luminous Infrared Galaxies (LIRGs; $\text{L}_{\text{IR}}\ge10^{11}\text{L}_{\odot}$) and Ultra-LIRGs (ULIRGs; $\text{L}_{\text{IR}}\ge10^{12}L_{\odot}$) are dominated by reprocessed thermal dust emission, due to a combination of  star-formation and AGN activity, with star-formation typically being the more dominant component \citep[e.g.\,][]{Watabe09,Elbaz10}. Locally, these sources are rare, although out to $z\sim1$ they become more numerous and increasingly dominate the infrared luminosity function of galaxies with increasing redshift \citep[e.g.\,][]{LeFloch05,Perez-Gonzalez05,Caputi07,Magnelli09,Rodighiero10,Eales10}.  (U)LIRGs are thought to trace a phase of intense star-formation activity, which is likely followed by, or partially concurrent with, an episode of vigorous black hole accretion. It is postulated that upon the cessation of these phases, each produces an early-type galaxy \citep{Genzel01,Farrah03, Lonsdale06,Veilleux09}.

Studies using the Multiband Imaging Photometer for {\it Spitzer} (MIPS; \citealt{Rieke04}) instrument onboard the {\it Spitzer Space Telescope} \citep{Werner04} have identified high-redshift ULIRGs from their $24\,\mu$m emission \citep[e.g.\,][]{Yan04,Houck05,Weedman06,Fiore08,Dey08,Farrah08,Fiore09}. \citet{Dey08} exploited this technique in the  Bo\"{o}tes field  of the NOAO Deep Wide Field Survey (NDWFS) and presented a sample of ULIRGs selected by the color cut $R-[24]\ge14$ (Vega magnitudes; $S_{24}/S_{R} \ge 1000$). Applying this selection scheme effectively identifies high-redshift infrared luminous galaxies containing large amounts of dust-obscuration, and which would be absent from UV-selected samples. This color selection preferentially identifies the rest-frame 7.7 $\mu$m polycyclic aromatic hydrocarbon (PAH) feature found in star-forming galaxies and causes the redshift distribution to have a biased average at $z\sim2$. Also, at $z\sim2$, the DOG selection falls within range of the power-law component of AGN emission in the mid-IR, which also identifies a population of active galactic nuclei (AGN). It is proposed that these dust-obscured galaxies (DOGs) are the latter stage of the sub-millimeter galaxy (SMGs; \citealp{Hughes98,Smail97,Barger98,Blain99}, among others) phase where an AGN is triggered while star formation is still occurring, causing some dust to be heated to higher temperatures \citep{Dey08} than in classic $850\,\mu$m selected SMGs. \citet{Pope08} found that  30\% of the SMGs are also DOGs, and of those SMG-DOGs, 30\% are AGN-dominated ($\ge50\%$ AGN contribution in mid-IR), consistent with this scenario. Using high resolution optical and near-IR imaging from the {\it Hubble Space Telescope} to investigate DOG morphology, the studies of \citet{Bussmann09b} and \citet{Bussmann11} found that the morphologies of bump (star-forming) DOGs, power-law (AGN dominated) DOGs, SMGs, and high redshift quiescent distant red galaxies (DRGs) are  consistent with the picture that major merger-driven systems eventually all evolve into compact relaxed passive galaxies ({\citealt{Springel05} and references therein}). Furthermore, \citet{Narayanan10} used N-body and hydrodynamic simulations to model the temporal evolution of high redshift galaxies and found that at the peak of the merger-driven galaxies' star formation rate, a galaxy can both be identified as an SMG and a DOG. The same study also found that during the stages after final coalescence, merger-driven DOGs transition from being star-formation dominated to being AGN dominated. \\
 
 The launch of the {\it Herschel Space Observatory}\footnotemark \footnotetext{{\it Herschel} is an ESA space observatory with science instruments provided by European-led Principal Investigator consortia and with important participation from NASA.} \citep{Pilbratt10} enables the direct observation of DOGs in the far-IR regime, instead of extrapolating from spectral energy distribution (SED) templates or stacking (e.g.\,\citealt{Dey08,Pope08}).\,\citet{Melbourne12} studied  {\it Herschel}-detected DOGs with spectroscopic redshifts and showed that DOGs classified by their near-IR SEDs as either bump (star-forming) or power-law (AGN-dominated) have $250\,\mu$m/$24\,\mu$m flux-density ratios that are consistent with local ULIRGs of the respective classes. \citet{Penner12} used {\it Herschel} data to show that DOGs' high rest-frame MIR/UV flux density ratios are due to varying amounts of UV dust obscuration, and speculated that it is caused by differing degrees of alignment between dust and stars, or simply by the differences in total dust content.\\

The focus of this paper is to extend the far-IR study of DOGs to a complete and statistically meaningful sample in order to accurately characterize their far-IR emission and calculate infrared luminosities. We generate our DOG catalog using Subaru $r^{+}$ band and MIPS $24\,\mu$m data from the Cosmological Evolution Survey (COSMOS; \citealt{Scoville07}) and combine it with multi-wavelength data in the far-IR from the {\it Herschel} Multi-tiered Extragalactic Survey\footnotemark \footnotetext{\textcolor{blue}{\hyperlink{link:hermes}{http://hermes.sussex.ac.uk/}}} (HerMES; \citealt{Oliver12}). We calculate IR luminosities, star formation rate (SFR) and dust temperatures for all DOGs detected at 250 $\mu$m and employ a stacking analysis to calculate the average properties of the undetected population in SPIRE and thus to probe fainter luminosities. For DOGs at $z=1.5-2.5$, we generate a luminosity function and calculate the star-formation rate density at $z\sim2$.    \\

This paper is organized as follows. Section~\ref{sec:data} describes the dataset and sample selection. The results and our analysis are presented in Section~\ref{sec:res}. We summarize our conclusions in Section~\ref{sec:conc}. Unless specifically stated, all magnitudes are reported in the AB system, where $-2.5\text{log}_{10}S_{\nu}(\mu\text{Jy}) + 23.9 = \text{AB mag}$, and assume a standard $\Lambda$CDM cosmology with $H_{0}=70\text{ km s}^{-1} \text{Mpc}^{-1}, \Omega_{\text{M}} = 0.3$, and $\Omega_{\Lambda} = 0.7$. 

%%%%%%%%%%%%%OBSERVATIONAL DATA AND SAMPLE SELECTION %%%%%%%%%%%%%%

\section{Data and Sample Selection}
\label{sec:data}

\subsection{Far-Infrared Data}

The 250, 350, and 500 $\mu$m far-IR data were obtained using the {\it Herschel}-Spectral and Photometric Imaging Receiver (SPIRE; \citealt{Griffin10,Swinyard10}) as part of HerMES, with an area coverage that completely overlaps with the MIPS observations of the 2\,deg$^{2}$ COSMOS field. We use the first data release (DR1) of  HerMES maps that were processed using the \texttt{smap} pipeline \citep{Levenson10}. The reduced maps reach  3$\sigma$ point source depths of 8, 10, and 14 mJy, in the 250, 350, and 500  $\mu$m channels respectively, where $\sigma$ is the combined instrumental and confusion noise. For sources with $S_{250} \ge 3\sigma$, we use the measured photometry from the HerMES cross-identification catalog (XID). This catalog uses known positions of $24\,\mu$m sources  as a prior, and estimates SPIRE fluxes via linear inversion methods. Model selections are used to account for, and prevent overfitting, and to optimize the $24\,\mu$m input. The fitting method is outlined in more detail in \citet{Roseboom10}.
 
\subsection{Optical and Mid-Infrared Data}
We use deep Subaru Suprime-Cam  \citep{Komiyama03} aperture-corrected $r^{+}$ photometry supplied by the COSMOS catalog \citep{Capak07}. The $5\sigma$ point-source depth for a $3''$ aperture is 26.8 mag. 

The near-IR data are from {\it Spitzer} observations carried out by the COSMOS {\it Spitzer} Survey (S-COSMOS; \citealt{Sanders07}) using the Infrared Array Camera (IRAC; \citealp{Fazio04}) and MIPS. The IRAC 5$\sigma$ depths at 3.6, 4.5, 5.6, and 8.0 $\mu$m for an aperture-corrected 1.9$''$ aperture, are 0.50, 0.6, 3, and 5 $\mu$Jy, respectively. The MIPS $24.0\,\mu$m 5$\sigma$ point source depth is $80\,\mu$Jy \citep{lefloch09}.

We next generate a MIPS\,$24\,\mu$m-selected catalog that combines the Subaru and {\it Spitzer} datasets, using a two-step cross-matching process within the 2 deg$^2$ of the Subaru deep area in order to find optical counterparts for each source \citep{fu10}. Firstly, the $24\,\mu$m coordinates are matched to the closest IRAC detection within a 2$''$ search radius, then the nearest optical counterpart is identified within 1$''$ of the IRAC position. Finally, sources near bright stars that were within the Subaru/optical and {\it Spitzer}/IRAC 3.6 $\mu$m coverage were removed from the catalog to avoid contamination. The final $24\,\mu$m catalog is $\ge90\%$\,complete above $S_{24}\ge80\,\mu$Jy and contains 28,639 sources with $S_{24}\ge100\,\mu$Jy.

 \subsection{Sample Selection}
\begin{figure}
\begin{minipage}[b]{\linewidth}
\centering
\includegraphics[width=\textwidth]{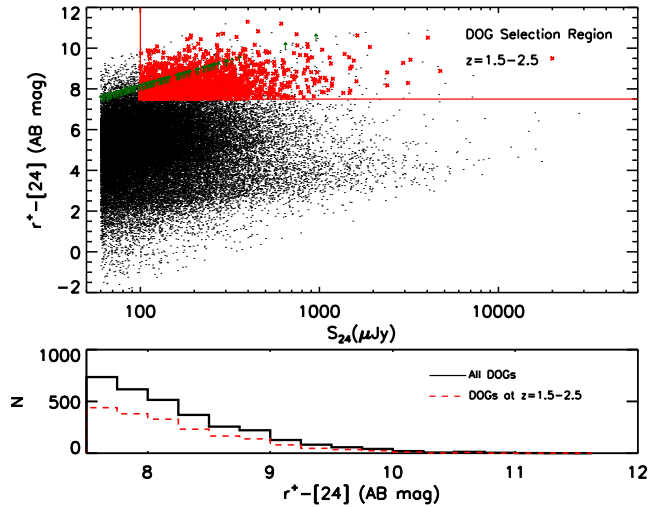}
\caption{$r^{+}$-[24] as a function of $24\,\mu$m flux ({\it top panel}) and $r^{+}$ - [24]  distribution for DOGs ({\it bottom panel}). DOGs are selected to have  $r^{+}$ - [24] $\ge$ 7.5 AB mag and $S_{24} \ge 100$ $\mu$Jy. DOGs  with $z=1.5-2.5$ are highlighted in red, while green arrows are lower limits for sources that were undetected in the $r^{+}$-band. The distribution of $r^{+}$ - [24] for all DOGs compared to those at $z=1.5-2.5$ shows that DOGs in this redshift range are not biased with respect to the full sample.}
\label{fig:R24vs24}
\end{minipage}
\end{figure}

 DOGs are selected in the standard manner, by identifying sources with $r^{+} - [24] \geq 7.5 $ (AB mag; $S_{24}/S_{r^{+}}\geq1000$) and we require $[24] \leq 18.90 \text{ mag}\text{ }(S_{24} \geq 100\mu$Jy) due to the depth of the $24\,\mu$m data. Using these criteria, 3,077 of the 28,639 (11\%) COSMOS $24\,\mu$m sources with $S_{24}\ge100\,\mu J$y are identified as DOGs (Fig.\,\ref{fig:R24vs24}). The near-IR SED of each DOG is examined using IRAC photometry ($\ge 5\sigma$) to classify whether a DOG contains  a bump-like feature or resembles a power-law.  For this study, a ``bump" DOG is defined if it satisfies one of the following: $S_{3.6} \le S_{4.5} \ge S_{8.0}$; $S_{4.5} \le S_{5.8}  \ge S_{8.0}$; or $S_{3.6}\le S_{4.5} \ge S_{5.8}$. Here $S_{[3.6, 4.5, 5.8, 8.0]}$ represent the flux densities in the 4 IRAC channels. Conversely, we label a DOG as ``power-law"  if it satisfies $S_{3.6} \le S_{4.5} \le S_{5.8} \le S_{8.0}$. Previous studies have interpreted sources that feature a bump in the near-IR SED to be the stellar continuum peak at rest-frame $1.6$ $\mu$m, tracing stellar emission and likely star-formation dominated (e.g.\,\citealp{Yan05,Sajina07}), while a power-law is dominated by AGN continuum emission (e.g.\,\citealp{Weedman06,Donley07}). Bump DOGs compose 47\% of our sample, while power-law DOGs are rarer,  totaling 10\%. The remaining 43\% are not classified due to one of two possibilities: insufficient or low signal to noise IRAC data; or an SED shape that does not satisfy the above criteria. For the latter case,  most of the sources are at $z < 2$ (median of $z=1.1$), such that the rest-frame $1.6\,\mu$m stellar continuum peak lies outside the wavelength range of the IRAC channels.

 \subsection{Redshifts} 
 All redshifts used in this paper are from COSMOS. Spectroscopic redshifts are used when available (35 sources, 1\%; \citealt{lilly07}, Kartaltepe et al., in prep), although virtually all of our DOG sample (2979 sources, 97\%) use photometric redshifts. The photometric redshifts are derived from 30 photometric bands \citep{Ilbert09}, providing $\sigma_{\Delta z/(1+z)} = 0.02$, for $24\,\mu$m sources that lie at $z=1.5-3$ and have the same $r^{+}$ mag range as DOGs. The 61 DOGs that are X-ray detected use photometric redshifts that also account for AGN flux variability \citep{Salvato09}. Also, two sources do not have a redshift estimate and are excluded from our sample. We note that the sharp peak in the redshift distribution at $z=1.95$ is due to rounding from the redshift values associated with the bin size used and no spatial correlation is observed. 

\begin{figure}
\begin{minipage}[b]{\linewidth}
\centering
\includegraphics[width=\textwidth]{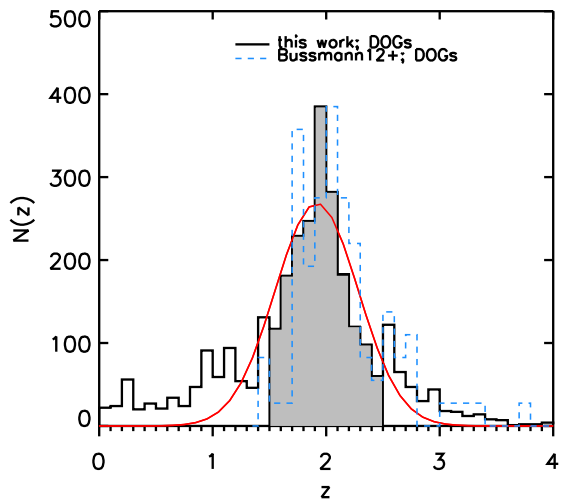}
\caption{Photometric redshift distribution of DOGs in the COSMOS field. We show the DOG distribution from \citet{Bussmann12} normalized to have equal peaks for comparison. The filled region highlights the range $z=1.5-2.5$, considered for our analyses in Sections~\ref{ssec:irlf} and~\ref{ssec:smbu}. We find $\langle z \rangle = 1.9\pm 0.3$, assuming a Gaussian distribution, as shown in red. } 
\label{fig:zhist}
\end{minipage}
\end{figure}

The redshift distribution of the final sample of 3075 DOGs is shown in Fig.~\ref{fig:zhist}, with a mean of $\langle z\rangle=1.9\pm0.3$. The sample of 90 DOGs in the  Bo\"{o}tes field with spectroscopic redshifts from \citet{Bussmann12}, normalized to have an equal peak with our sample, is also shown. The two samples have a consistent mean $z$ of $1.9\pm0.02$ and $2.1\pm0.5$, for our sample and the \citet{Bussmann12} sample, respectively. 

%%%%%%%%%%%%%%%%%%%%%%%%%
% FIgure: R-[24] vs S_24 
%%%%%%%%%%%%%%%%%%%%%%%%%

\section{Analyses and Results}
\label{sec:res}
%%%%%%%%%%%%%%%%%%%%%%%%%
% FIgure: N(z) vs. z
%%%%%%%%%%%%%%%%%%%%%%%%%
%%%%%%%%%%%%%%%%%%%%%%%%%

\subsection{Far-Infrared Spectral Energy Distributions}
\label{sec:seds}

Using the COSMOS redshifts and {\it Herschel} 250, 350 and 500 $\mu$m photometry, we fit the far-IR SED and calculate the rest-frame IR luminosity ($8-1000$ $\mu$m) and characteristic dust temperature.  We divide the DOGs into two subsamples based on $250\,\mu$m detections because this SPIRE channel offers the deepest far-IR observations and the smallest beam size. A DOG is considered $\it Herschel$-detected if it satisfies $S_{250} \ge 3\sigma_{250}$ (where $\sigma_{250}$ is the total uncertainty due to the instrumental and confusion noise), and undetected otherwise. Of our DOG sample, 51\% are thus {\it Herschel}-detected. To calculate the characteristic dust temperature, for each of these we use the available SPIRE flux densities to fit a modified blackbody of the form
\begin{equation}
S_{\nu} \propto B_{\nu}(T_{\text{dust}})\nu^{\beta},
\label{eqn:modbb}
\end{equation}
where $\nu$ is frequency, $\beta$ is the dust emissivity, fixed to the typical value of 1.5 \citep{Draine03}, $T_{\text{dust}}$ is the dust temperature and $B_{\nu}$ is the Planck function, defined as
\begin{equation}
B_{\nu} = \frac{2h\nu^3}{c^2}\frac{1}{e^{h\nu/k_{\text{B}}T_{\text{dust}}}-1}.
\end{equation}
 Here $h$ is Planck's constant, $c$ is speed of light, and $k_{\text{B}}$ is Boltzmann's constant. The temperature we calculate is insensitive to and consistent with the reported error bars from varying $\beta$ slightly. All $250\,\mu$m-detected sources have measured flux densities  at either $350\,\mu$m or $500\,\mu$m, although it is not required to satisfy the $3\sigma$ limit (including confusion) in these wavelengths. For the sources that have low significance detections ($\le 3\sigma$) at $350\,\mu$m and/or $500\,\mu$m, we allow the full range of the uncertainties in flux densities when fitting for their IR luminosities and dust temperatures.
 
\begin{figure}
\begin{minipage}[b]{\linewidth}
\centering
\includegraphics[width=\textwidth]{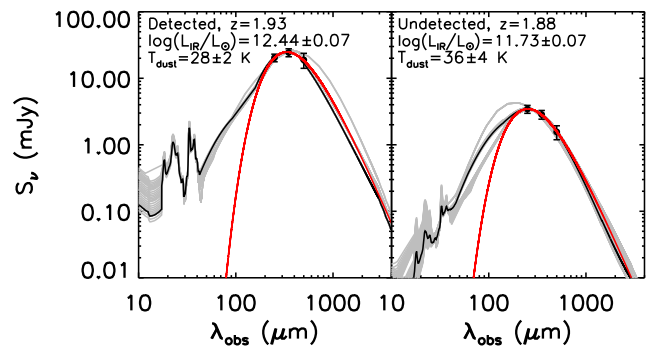}
\caption{Example SED fitting for a {\it Herschel}-detected DOG ({\it left panel}) and an undetected DOG ({\it right panel}). The black curve shows the best fitting template to the SPIRE data points (black circles) and the gray curves show CE01 templates that provide acceptable fits consistent with the error bars.  The red curve shows the best fit modified blackbody, which we use to calculate the dust temperature.  } 
\label{fig:sed}
\end{minipage}
\end{figure}

We derive estimates of the IR luminosity by fitting the available SPIRE data to the SED template library of \citet{Chary01} (hereafter CE01). The template with the minimum $\chi^{2}$ is chosen for the best fit. The uncertainty in IR luminosity is derived by first producing 1000 mock catalogs for each source that assume a Gaussian distribution centered around the measured SPIRE flux density, with a dispersion equal to the average flux density error. The IR luminosity per source is recalculated 1000 times and the standard deviation of the IR luminosity distribution is the error in our calculation. Examples of the SED template and modified blackbody fitting are shown in Fig.~\ref{fig:sed}.

The IR luminosity ($8-1000\,\mu$m) is converted to star formation rate using \citep{Kennicutt98}  
 \begin{equation}
 \text{SFR} (\text{M}_{\odot}\text{yr}^{-1})=1.72 \times10^{-10}L_{\rm{IR}} (L_{\odot}),
 \label{eq:sfr}
 \end{equation} 
which assumes a Salpeter initial mass function (IMF). We note that in our study we assume that UV emission will provide negligible contribution to the total star-formation rate, as validated by \citet{Penner12}.

%%%%%%%%%%%%%%%%%%%%%%%%%
% FIgure:DOG SED
%%%%%%%%%%%%%%%%%%%%%%%%%
%%%%%%%%%%%%%%%%%%%%%%%%%

To measure the average flux density of the {\it Herschel}-undetected DOGs, we bin the sources in redshift and for each bin stack on the SPIRE residual maps. These maps are generated by performing a blind extraction and PSF-subtraction to prevent contamination of individually detected sources.  We use the publicly available {\sc {idl}}  stacking library from \citet{Bethermin10} to perform the stacking\footnotemark \footnotetext{The {\sc{idl}} stacking library from \citet{Bethermin10} is available at \textcolor{blue}{\hyperlink{link:stack}{http://www.ias.u-psud.fr/irgalaxies/downloads.php}}}. Each stacked image was converted from the native $\text{Jy} \text{ beam}^{-1}$ to $\text{Jy} \text{ pixel}^{-1}$ and aperture photometry with an aperture size equal to $22''$, $30''$, and $42''$ for 250, 350 and 500 $\mu$m respectively, is performed to calculate the flux of the stacked images. These aperture flux densities are consistent with those measured in the central pixel when the stacked map is in units of Jy~beam$^{-1}$. 

The observed stacked flux densities are corrected for the boosting from clustering bias by dividing by factors of 1.07, 1.10 and 1.20 at 250, 350 and 500\,$\mu$m, respectively. The appropriate correction factors vary with clustering strength and are thus population dependent. These values were calculated by \citet{Bethermin12} for $24\,\mu$m sources and are valid for DOGs because the observed correlation lengths, $r_{0}$ (a proxy for clustering amplitude), for DOGs \citep{Brodwin08} and the parent population of $24\,\mu$m sources \citep{Magliocchetti08, Starikova12} are consistent. Errors in the photometry are calculated from bootstrapping the sources to be stacked. For each redshift bin, the clustering-corrected SPIRE flux densities of {\it Herschel}-undetected DOGs are set to equal the median stacked flux densities and the IR luminosity and dust temperature are calculated using the same method as for the Herschel-detected DOGs. The (clustering-corrected) stacked fluxes and errors, and the resulting average infrared luminosities and dust temperatures are shown in Table~\ref{tab:stack}.

\begin{figure}
\begin{minipage}[b]{\linewidth}
\centering
\includegraphics[width=\textwidth]{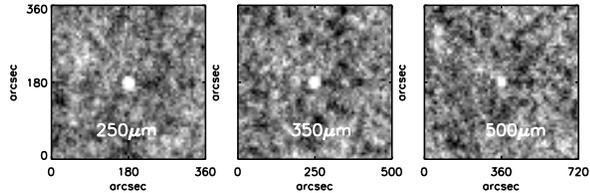}
\caption{Example median stacking results of $\it Herschel$-undetected DOGs at 250, 350, and $500\,\mu$m for undetected DOGs at $z=1.75$ to $2.00$}.
\label{fig:stacked}
\end{minipage}
\end{figure}

We note that the average stacked $250\,\mu$m flux density for the {\it Herschel}-undetected DOGs is $4.1\pm0.7$ mJy, which is a factor of 2 lower than the $250\,\mu$m catalog detection limit. In Fig.~\ref{fig:stacked} we show an example of the median stacked images for 250, 350 and $500\,\mu$m from left to right at $z=1.75 - 2.00$ and an example SED using stacked SPIRE flux densities for an {\it Herschel}-undetected DOG at $z=1.88$ is shown in the right panel of Fig.~\ref{fig:sed}. Each image stack is large enough to provide a good estimate for the background noise. 

%%%%%%%%%%%%%%%%%%%%%%%%%
% Figure: Stacked SPIRE images
%%%%%%%%%%%%%%%%%%%%%%%%%
%%%%%%%%%%%%%%%%%%%%%%%%%

%%%%%%%%%%%%%%%%%%%%%%%%%
%Table: Dust Temperature and IR Luminosities
%%%%%%%%%%%%%%%%%%%%%%%%%
 \begin{table*}[ht]
 \centering
 \begin{minipage}{12cm}
\caption{SPIRE Stacking Results} % title of Table
\centering  % used for centering table
\tabcolsep=0.10cm
\begin{tabular}{c c c c c c c c} % centered columns (4 columns)
\hline\hline                        %inserts double horizontal lines
bin (n)$^{a}$ & $z$   & $S_{250}^{b}$ & $S_{350}^{b}$ &  $S_{500}^{b}$ & $N^{c} $  &  $L_{\text{IR}}^d$  &  $T_{\text{dust}}^d$ \\[0.5ex]
  &        &  (mJy)  & (mJy) & (mJy) &  & ($\times 10^{12}\,L_{\odot}$) &  (K)\\[0.5ex]

%heading
\hline                  % inserts single horizontal line
1 &$<1.5$ & $4.2 \pm 0.5 $ & $3.0 \pm 0.4$ & $2.1 \pm 0.4$ & $354$  & $0.16\pm0.13$ & $25.1\pm5.5$\\ % inserting body of the table
2&$1.50 - 1.75$ & $3.5 \pm 0.5 $ & $2.6 \pm 0.5$ & $1.7 \pm 0.4$ & $218$ & $0.37\pm0.02$ & $34.6\pm0.9$\\
3&$1.75 - 2.00$ & $3.4 \pm 0.5 $ & $2.9 \pm 0.4$ & $1.6 \pm 0.4$ & $406$ & $0.52\pm0.07$ & $37.2\pm0.9$\\
4&$2.00 - 2.25$ & $4.3 \pm 0.6 $ & $3.9 \pm 0.6$ & $1.7 \pm 0.5$ & $237$ & $0.68\pm0.06$ & $40.0\pm1.0$\\
5&$2.25 - 2.50$ & $4.3 \pm 1.1 $ & $3.7 \pm 0.9$ & $2.7 \pm 0.7$ & $82$   & $1.00\pm0.21$ & $37.8\pm0.8$\\
6&$>2.5$ & $4.9 \pm 0.7 $ & $4.0 \pm 0.6$ & $2.9 \pm 0.6$ & $185$ & $1.71\pm2.8$ & $44.3\pm6.3$ \\ % [1ex] adds vertical space
\hline %inserts single line
\end{tabular}
\footnotetext{{\bf Notes}}
\footnotetext{$^{a}$ Bin number used to label stacked undetected DOGs in fig.~\ref{fig:tdvslfir}}
\footnotetext{$^{b}$ Measured flux densities are from median stacking. The errors are from bootstrapping.}
\footnotetext{$^{c}$ Number of sources per redshift bin.}
\footnotetext{$^{d}$ Average and standard deviation per bin.}
\label{tab:stack} % is used to refer this table in the text
\end{minipage}
\end{table*}

%%%%%%%%%%%%%%%%%%%%%%%%%

\begin{figure}
\begin{minipage}[b]{\linewidth}
\centering
\includegraphics[width=\textwidth]{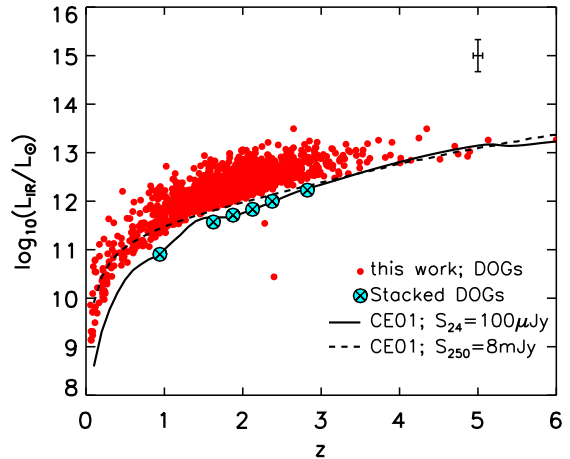}
\caption{IR luminosity as a function of redshift for {\it Herschel}-detected DOGs and the median IR luminosity for stacked DOGs. A representative template from \citet{Chary01} scaled to the DOG $24\,\mu$m  (solid curve) and  250 $\mu$m detection limit (dashed curve) are also shown. A typical error bar is shown at the top right. The apparent trend that IR luminosity increases with redshift is a selection effect. }
\label{fig:lfirvsz}
\end{minipage}
\end{figure}

%LFIR vs. z
Fig.~\ref{fig:lfirvsz} shows IR luminosities of the {\it Herschel}-detected DOGs as a function of redshift. The average IR luminosity for {\it Herschel}-detected and undetected DOGs is $(2.8 \pm 0.3 )\times10^{12}\,L_{\odot}$ and  $(6.0 \pm 1.0)\times10^{11}\,L_{\odot}$, respectively. LIRGs ($10^{11} \le L_{\text{IR}}\,(L_{\odot}) \le 10^{12}$) comprise 15\% of {\it Herschel}-detected DOGs and 75\% for {\it Herschel}-undetected DOGs. ULIRGs ($10^{12} \le L_{\text{IR}}\,(L_{\odot}) \le 10^{13}$) make up 78\% of the {\it Herschel}-detected and 15\% for {\it Herschel}-undetected DOGs.\,\,Hyper-luminous infrared galaxies (HLIRGs ($\ge 10^{13}\,L_{\odot}$) are the rarest, totaling 2\% for {\it Herschel}-detected DOGs and none for {\it Herschel}-undetected DOGs. Although we note that there is additional uncertainty in the fractional contributions of the undetected sources, due to the use of stacked average fluxes, which minimizes the contribution from extreme sources. {\it Herschel}-detected power-law, or AGN-dominated DOGs, have on average $L_{\text{IR}} = (4.5\pm0.5)\times10^{12}\,L_{\odot}$, making them more IR-luminous than {\it Herschel}-detected bump, star-forming DOGs, which have $L_{\text{IR}} = (3.1\pm0.4 )\times10^{12}\,L_{\odot}$, which is consistent with the findings of \citet{Melbourne12}. 
Selection effects are investigated by calculating the IR luminosity of a representative CE01 template, scaled such that $S_{24} = 100\,\mu$Jy or $S_{250} = 8$\,mJy, as shown in Fig.~\ref{fig:lfirvsz}. The local maxima in IR luminosity at $z\sim1.5$ in the $24\,\mu$m curve is due to the rest-frame $9.7\,\mu$m silicate absorption feature. The lack of DOGs below the $24\,\mu$m and $250\,\mu$m limit at a given redshift leads us to conclude that the apparent trend in Fig.~\ref{fig:lfirvsz} that IR luminosity is increasing with redshift is a selection effect.

 %%%%%%%%%%%%%%%%%%%%%%%%%
% Figure: IR Luminosity vs. redshift
%%%%%%%%%%%%%%%%%%%%%%%%%
%%%%%%%%%%%%%%%%%%%%%%%%%

\begin{figure*}
%\begin{minipage}[b]{\linewidth}
\centering
\includegraphics[width=\textwidth]{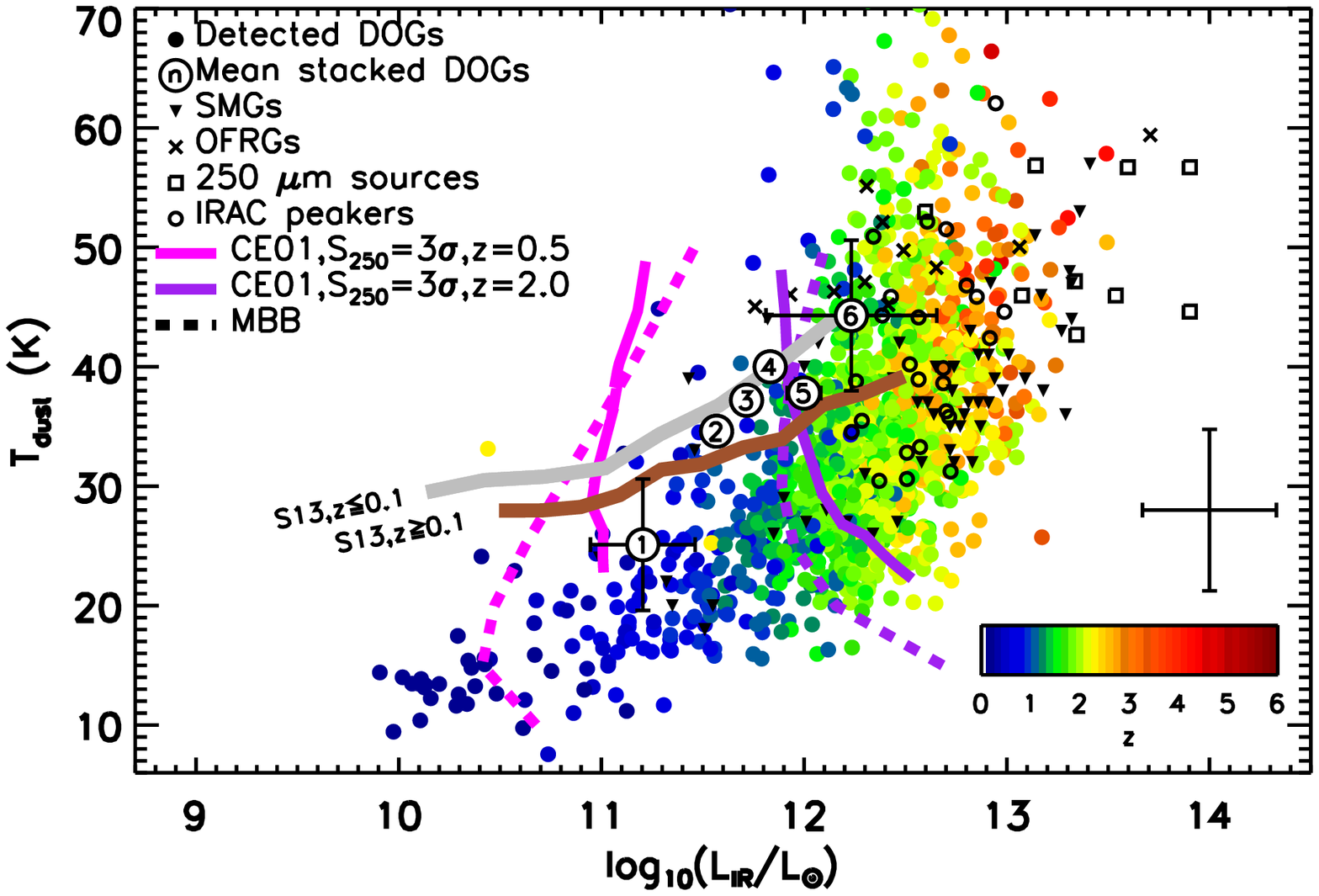}
\caption{Dust temperatures and infrared luminosities for DOGs compared to other $z\sim2$ galaxy populations (SMGs:\,\citealt{Magnelli12},\,\,OFRGs:\,\citealt{Casey09,Magnelli10},\,\,$250\,\mu$m sources:\,\citealt{Casey11},\,\,IRAC peakers:\,\citealt{Magdis10}). We show {\it Herschel}-detected DOGs, colored by redshift. The mean dust temperature and IR luminosity per redshift bin for the stacked DOGs are labeled by their bin number from table~\ref{tab:stack}, with their dispersions shown as error bars. We note that some of the bins have dispersions are too small to display and thus appear to be invisible.The magenta and purple solid curves are generated from estimating the dust temperature and calculating IR luminosity from \citet{Chary01} templates at $z = 0.5$ and $z=2.0$, respectively, with $S_{250}=8$\,mJy, our $3\sigma$ detection limit. The dashed curves are generated from calculating the IR luminosities of equation~\ref{eqn:modbb} at fixed temperatures, also analyzed at $z=0.5$ and $z=2$. The gray and brown curves are the  $T_{\rm{dust}}-L_{\rm{IR}}$ relations for $z\le0.1$ and $z\ge0.1$, respectively, from \citet{Symeonidis13}. We conclude that the apparent trend that hotter sources are at higher redshifts is caused by the $T_{\rm{dust}}-L_{\rm{IR}}$ relation and the redshift dependent selection limit in $L_{\rm{IR}}$. The absence of warm, low luminosity (low redshift) sources is a selection effect. However, cool, high luminosity sources would be detected in our data, and the dearth of these sources is not a selection effect.} 
\label{fig:tdvslfir}
%\end{minipage}
\end{figure*}

%Tdust vs. LFIR
Figure~\ref{fig:tdvslfir} shows dust temperature as a function of IR luminosity for DOGs, color-coded by redshift. The average characteristic dust temperature is  $T_{\text{dust}} = (34 \pm 7)$\,K and $(37\pm5)$\,K for {\it Herschel}-detected and undetected DOGs, respectively. {\it Herschel}-detected power-law DOGs and bump DOGs have average $T_{\text{dust}} = (37\pm6)$\,K and $(35\pm7)$\,K,respectively, which is consistent with each other. We investigate sample selection effects in the $T_{\text{dust}} - L_{\rm{IR}}$ plane by considering both the IR luminosity of fixed temperature modified blackbody SEDs (equation~\ref{eqn:modbb}) and the CE01 templates at $z=0.5$ and $z=2.0$ for $S_{250} = 8$ mJy (fig.~\ref{fig:tdvslfir}). The temperatures of the CE01 SEDs are calculated by fitting the template 250, 350, and $500\,\mu$m flux densities with a modified blackbody as in our data. Figure~\ref{fig:tdvslfir} also confirms that the $250\,\mu$m flux density limit biases against lower luminosity sources and the luminosity limit is a function of redshift (see also Fig.~\ref{fig:lfirvsz}). Furthermore, the apparent correlation between $T_{\text{dust}}$ and $z$ is in fact caused by a combination of the redshift-dependent $L_{\rm{IR}}$ selection limit and the correlation between $L_{\rm{IR}}$ and $T_{\text{dust}}$ \citep[e.g.\ ][]{Symeonidis13,Hwang10}. The selection limits at $z=0.5$ on Fig.~\ref{fig:tdvslfir} also show that at low redshift ($z\lesssim1$) there is a bias against the warmer sources, which results in an apparent difference in the dust temperature distribution of {\it Herschel}-detected DOGs compared to the observed relationship locally and at higher redshifts \citep[e.g.\ ][]{Chapman04, Symeonidis13,Hwang10}. The $24\,\mu$m flux density limit produces a similar effect as the $250\,\mu$m limit.

The mean stacked IR luminosities and dust temperatures, per redshift bin, of the {\it Herschel}-undetected DOG population are displayed on Fig.~\ref{fig:tdvslfir} and are less sensitive to these selection biases. The lowest redshift bin (bin 1, $z\le1.5$) is offset relative to the other redshift bins because it covers a wide redshift interval in which the IR luminosity limit has a steep slope (Fig.~\ref{fig:lfirvsz}).  The dearth of sources at high luminosities and low dust temperatures is not a selection effect as these sources would have been detected by our data. This is consistent with \citet{Symeonidis13}, who found that cold cirrus-dominated SEDs \citep{Rowan-Robinson10} are rare in the most IR luminous galaxies.

 %%%%%%%%%%%%%%%%%%%%%%%%%
% Figure: Dust Temperature vs. IR Luminosity
%%%%%%%%%%%%%%%%%%%%%%%%%
%%%%%%%%%%%%%%%%%%%%%%%%%%%%%%%
%%%%%%% Compare TD and IR luminosity to other galaxy populations

 %%%%%%%%%%%%%%%%%%%%%%%%%
%Table: Dust Temperature and IR Luminosities
%%%%%%%%%%%%%%%%%%%%%%%%%
  \begin{table}
\caption{Average DOG IR Luminosities and Dust Temperatures} % title of Table
\centering  % used for centering table
\begin{tabular}{c c c c} % centered columns (4 columns)
\hline\hline                        %inserts double horizontal lines
Type   & $L_{\text{IR}}$ & $T_{\text{dust}}$  \\ [0.5ex]% inserts table 
          &  $(\times10^{12}\,L_{\odot})$ & (K)\\[0.5ex]
%heading
\hline                  % inserts single horizontal line
Detected$^{\alpha}$ & $ 2.8 \pm 0.3 $ & $33\pm7$ \\ % inserting body of the table
Undetected$^{\alpha}$ & $0.6 \pm 0.1$ & $37\pm5$ \\
$\text{Bump}^{*}$ & $4.5\pm0.4$  & $34\pm7$  \\
$\text{Power-law}^{*}$ & $3.1\pm0.4$ & $37\pm6$ \\    % [1ex] adds vertical space
\hline %inserts single line
\end{tabular}
\footnotetext{{\bf Notes}}
\footnotetext{$^{\alpha}$ Detected sources satisfy: $S_{250}\ge3\sigma$;  undetected sources have $S_{250}<3\sigma$.}
\footnotetext{$^{*}$ Power-law (AGN-dominated) and bump (star-forming) median IR luminosities are from the detected sample.}
\label{tab:lirtd} % is used to refer this table in the text
\end{table}
%%%%%%%%%%%%%%%%%%%%%%%%%

 %%%%%%%%%%%%%%%%%%%%%%%%%
%Figure and Table: DOG IR Luminosity Function at z=2
%%%%%%%%%%%%%%%%%%%%%%%%%
\begin{figure*}
\centering
\includegraphics[width=\textwidth]{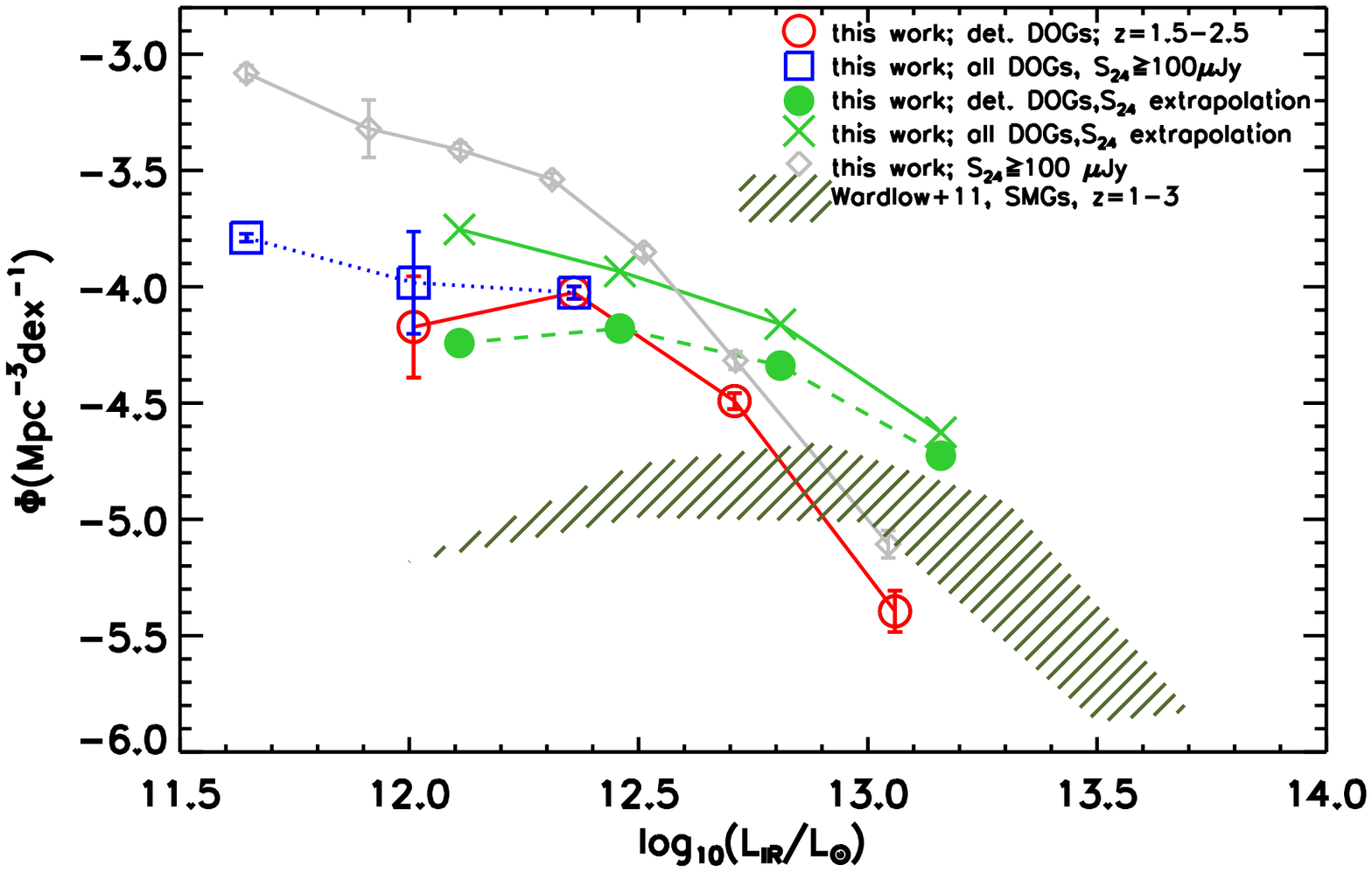}
\caption{IR luminosity function of DOGs in the COSMOS field at $z=1.5-2.5$ with $S_{24}\ge100\,\mu$Jy. Individually $\it Herschel$-detected DOGs and the results from stacking undetected DOGs are shown. We compare this to an IR luminosity function for {\it Herschel}-detected DOGs and all DOGs generated from $24\,\mu$m extrapolation using templates from \citet{Chary01}, classical SMGs \citep{Wardlow11}, and $24\,\mu$m selected galaxies with $S_{24}\ge100$ $\mu$Jy. The results of stacking allow us to estimate the faint end of the LF and we note that using $24\,\mu$m flux density to calculate IR luminosity results in overestimation. DOGs have a higher normalization,$\Phi^{\ast}$, but a lower luminosity turnover, $L^{\ast}$, than SMGs.}
\label{fig:doglf}
\end{figure*}

The results shown in Fig.~\ref{fig:tdvslfir} suggest that $z\sim2$ DOGs span a wider range of dust temperatures than $z\sim2$ SMGs (by which we mean $850\,\mu$m or 1\,mm selected sources) due to the different selection effects associated with each galaxy population. SMGs are biased towards detecting cold-dust dominated sources ($T_{\rm{dust}}\sim30-40$ K) because hot sources are missed by sub-mm surveys \citep{Chapman04,Casey09,Chapin09}. The discovery of optically-faint radio galaxies \citep[OFRGs;\ ][]{Chapman04,Casey09,Magnelli10} at $z\sim2$, which have similar stellar masses, radio luminosities, and UV spectra as SMGs, but have $T_{\rm{dust}}\sim40-60$ K, demonstrate this, while we also note that the radio-detection limit is biased against the coldest sources \citep[e.g.\,][]{Wardlow11}. DOGs are more insensitive to these selection biases and thus show a wider range of temperatures at $z\sim2$. \citet{Magdis10} found similar results when investigating the characteristic dust temperatures for IRAC peakers and showed that mid-IR selected sources bridge the gap in temperature ranges between OFRGs and SMGs. We note that the $250\,\mu$m selected sources suffer from the same selection biases as our {\it Herschel}-detected DOGs but shifted to higher luminosities due to their shallower $250\,\mu$m detection limit. 
 
\subsection{Infrared Luminosity Function at $z\sim2$}
\label{ssec:irlf}
 We compute  the IR luminosity function of DOGs using the 1/$V_{\rm{max}}$ method \citep{Schmidt78}, defined as

\begin{equation}
\Phi(L)\Delta L = \displaystyle\sum_{i}\frac{1}{V_{\text{max},\,i}},
\end{equation}
where $V_{\rm{max}}$ is the maximum comoving volume of the $i$th source such that it would be detected and included in the sample. We consider the peak of the redshift distribution using only DOGs at $z=1.5-2.5$. For the {\it Herschel}-detected DOGs we use two flux limits to determine $V_{\rm{max}}$: $S_{24} = 100\,\mu$Jy; and $S_{250} = 8$\,mJy. These are the two detection limits of the survey. For the {\it Herschel}-undetected sample, the $24\,\mu$m flux limit alone was used to calculate $V_{\rm{max}}$, as all redshift bins are detected in the stacks. We then calculate each {\it Herschel}-undetected DOGs' IR luminosity and contribution to the luminosity function using its redshift and the relevant stacked flux densities. The uncertainties are from Poisson statistics and binning errors, where the binning errors are calculated by generating the IR luminosity function 1000 times from IR luminosities calculated from artificial SPIRE flux densities described in Section~\ref{sec:seds} and taking the standard deviation per IR luminosity bin. The DOG IR luminosity function at $z\sim2$ is presented in Fig. \ref{fig:doglf} and Table \ref{tab:lf}. The faint end of the IR luminosity function for {\it Herschel}-detected and undetected DOGs are coadded, which affects the lowest luminosity bin for {\it Herschel}-detected DOGs the most, showing a 0.20 dex increase.
 
For comparison, the DOG IR luminosity function for {\it Herschel}-detected DOGs and all DOGs, calculated by extrapolating the infrared luminosity from  $S_{24}$ using CE01 templates, is  also shown in Fig.~\ref{fig:doglf}.  We find that the IR luminosities using this method are overestimated by a median factor of 1.8, consistent with the previous studies of $24\,\mu$m-selected galaxies at  $z\sim2$ \citep{Houck05,Yan07,Daddi07,Papovich07, Pope08, Nordon10, Nordon12,Elbaz10,Elbaz11,Magnelli11}, and affects both the shape and normalization of the IR luminosity function.

 %%%%%%%%%%%%%%%%%%%%%%%%%
%Figure: SFRD vs. z for DOGs
%%%%%%%%%%%%%%%%%%%%%%%%%
\begin{figure*}
\includegraphics[width=\textwidth]{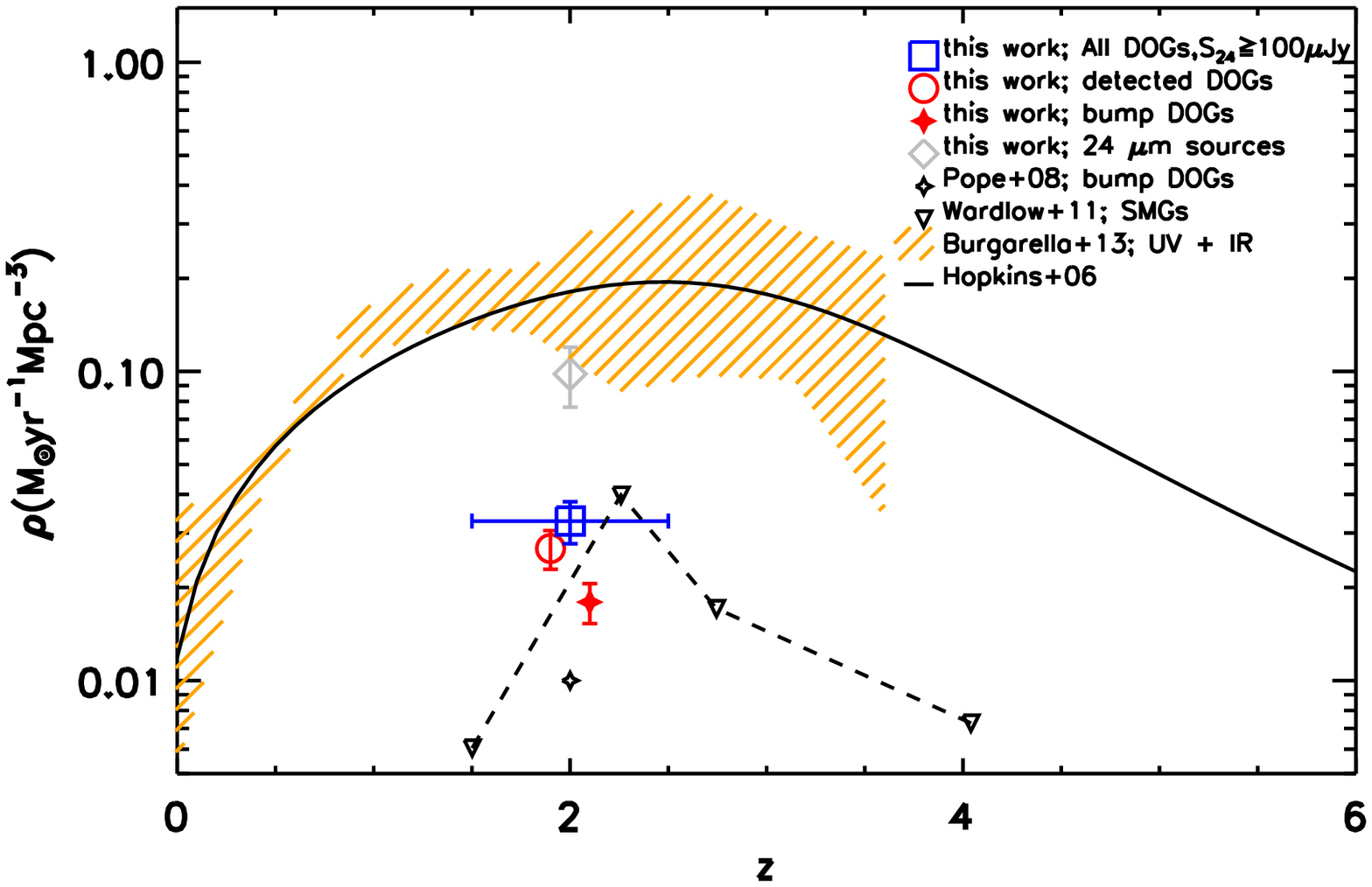}
\caption{Star formation rate density ($\rho_{\rm{SFR}}$) of all DOGs with $S_{24}\ge100\,\mu$Jy, {\it Herschel}-detected DOGs (offset by $z=-0.1$, for clarity), bump DOGs (offset by $z=+0.1$, for clarity) and all $S_{24}\ge100\,\mu$Jy sources in COSMOS at $z=1.5-2.5$.  We also show $\rho_{\rm{SFR}}$ for bump DOGs ($S_{24}\ge 100$ $\mu$Jy) at $z=2$ in the GOODS field from \citet{Pope08} and SMGs from \citet{Wardlow11} The evolution of $\rho_{\rm{SFR}}$ as a function of redshift from \citet{Hopkins06} and \citet{Burgarella13} are also shown. Based on these models, DOGs contribute $12-29\%$ to the total $\rho_{\rm{SFR}}$ of the Universe at $z\sim2$.}
\label{fig:sfrdvsz}
\end{figure*}
%%%%%%%%%%%%%%%%%%%%%%%%%%%%%%

We compare the number densities of DOGs to the parent population of sources with $S_{24}\ge100\,\mu$Jy (Fig.~\ref{fig:doglf}). The luminosity function of  $24\,\mu$m sources is calculated self-consistently using SPIRE data, including stacking on the {\it Herschel}-undetected galaxies. There are 5,932 sources in COSMOS with $S_{24} \ge 100\,\mu$Jy at $z=1.5-2.5$, of which 32\% are DOGs. Figure~\ref{fig:doglf} shows that DOGs have a smaller overall normalization in their luminosity function (since they are fewer in number) and their relative contribution to the  $24\,\mu$m number density increases with luminosity in agreement with the IR luminosity distribution of DOGs compared to $24\,\mu$m sources in \citet{Penner12}.

Figure~\ref{fig:doglf} also shows the $z=1-3$ SMG luminosity function by combining the $z=1-2$ and $z=2-3$ redshift bins from \citet{Wardlow11} for comparison of the distinct, but overlapping DOG and SMG \citep{Pope08} populations. DOGs are more common than SMGs at $z\sim2$, which is reflected in the higher normalization in the luminosity function, although in the HLIRG regime, SMGs dominate. This is consistent with the picture in which DOGs represent an evolutionary stage towards the end of the peak of star-formation rate, observed as the SMG phase \citep{Narayanan10}, they must have lower IR luminosities and star-formation rates on average. In this case, the relative scaling of the two luminosity functions indicate that the DOG phase is longer lived than the SMG phase. As is shown in Fig.~\ref{fig:sfrdvsz}, the total star-formation rate density ($\rho_{\rm{SFR}}$) provided by the two populations are approximately even despite the number and intensity of sources.

 %%%%%%%%%%%%%%%%%%%%%%%%%%%%
 \begin{table}
\caption{The IR Luminosity Function for DOGs at $z=1.5-2.5$} % title of Table
\centering  % used for centering table
\begin{tabular}{c c c c} % centered columns (4 columns)
\hline\hline                        %inserts double horizontal lines
$\text{log}_{10}(L_{\text{IR}}/L_{\odot})$ & $\Phi $ & $N^a$ \\ [0.5ex]% inserts table 
  &$(\text{Mpc}^{-3}\text{dex}^{-1})  $& \\
%heading
\hline                  % inserts single horizontal line
$11.40-11.80$ & $(-3.79\pm0.02)^b$ & $(660)^b$ \\
$11.80-12.15$ & $-4.17\pm0.21\,(-3.98\pm0.22)^b$ & $150\,(433)^b$  \\ % inserting body of the table
$12.15-12.50$ & $-4.03\pm0.03\,(-4.02\pm0.09)^b$ & $522$  \\
$12.50-12.85$ & $-4.49\pm0.03$ & 252 \\
$12.85-13.20$ & $-5.40\pm0.09 $ & 31 \\    % [1ex] adds vertical space
\hline %inserts single line
\end{tabular}
\footnotetext{{\bf Notes}}
\footnotetext{$^a$ $N$ is the number of sources per luminosity bin.}
\footnotetext{$^b$ Numerical values in parenthesis include the stacked contribution.}
\label{tab:lf} % is used to refer this table in the text
\end{table}

To calculate the contribution of DOGs with $S_{24} \ge 100\,\mu$Jy to the $\rho_{\rm{SFR}}$ of the Universe at $z\sim2$, we integrate the IR luminosity function and use equation~\ref{eq:sfr}. Figure~\ref{fig:sfrdvsz} shows DOGs compared to other $z\sim2$ galaxy populations. The total uncertainty in $\rho_{\rm{SFR}}$ is calculated from the quadrature sum of individual star-formation rate uncertainties and the standard deviation of $\rho_{\rm{SFR}}$ from the mock catalogs discussed in Section~\ref{sec:seds}. Horizontal error bars represent the considered redshift interval. The value of $\rho_{\rm{SFR}}$ for DOGs at $z=1.5-2.5$ is $(3.2\pm0.5)\times10^{-2} \text{ M}_{\odot}\text{yr}^{-1}\text{Mpc}^{-3}$ which contributes to $12-29\%$ of the overall $\rho_{\rm{SFR}}$ at $z=2$ calculated from UV and IR data shown by \citet{Hopkins06} and \citet{Burgarella13}. When comparing to $z=1.5-2.5$ sources with $S_{24}\ge100\,\mu$Jy, DOGs contribute 33$\%$ to the $24\,\mu$m $\rho_{\rm{SFR}}$. 

The {\it Herschel}-undetected and power-law sources provide non-dominant contributions to the total $\rho_{\rm{SFR}}$ of DOGs. The {\it Herschel}-undetected DOGs contribute 18\% and power-law DOGs contribute just 9\%. We note that even though power-law DOGs are thought to be dominated by AGN emission in the IRAC bands, their far-IR emission is still likely dominated by star-formation, as is the case for other far-IR luminous samples containing AGN \citep{Elbaz10,Lutz08,Lutz10}. Indeed, even studies of the most active AGN have revealed that SED fits for {\it Herschel}-detected AGNs always required a starburst component in order to appear bright in the far-IR \citep{Hatziminaoglou10}. As an attempt to quantify this claim, we use a simplified method to calculate an upper limit on the AGN contribution to the IR luminosity and star-formation rate in power-law DOGs and hence the contamination of $\rho_{\rm{SFR}}$ by AGN. We begin by scaling the AGN SEDs from \citet{Kirkpatrick12} to the $24\,\mu$m flux density of each power-law DOG and calculate the luminosity from the warm dust component. Then, by assuming that the warm dust component is entirely AGN-dominated and the cold dust component is entirely star-formation dominated, we can subtract the warm IR luminosity from the CE01 IR luminosity to calculate the residual contribution from star-formation. We find that power-law  DOGs each have a maximum average contribution of ~70\% to the IR luminosity, which could contaminate $\rho_{\rm{SFR}} $ by $\sim 0.2 \times 10^{-2} \text{ M}_{\odot}\text{yr}^{-1}\text{Mpc}^{-3}$, which is only 6\% of the total DOG $\rho_{\rm{SFR}}$. In addition, we also estimate the dispersion of AGN contribution by normalizing quasar SED templates from \citet{Elvis94}, \citet{Richards06}, \citet{Polletta07} and \citet{Dai12} to the average power-law DOG $24\,\mu$m flux density at $z=1.5-2.5$ and assume that the SEDs have no emission associated with star-formation. Under this assumption, the average AGN contributions  to the individual galaxies' IR luminosities range from 5\% to 65\%, depending on the SED, which corresponds to 0.005\% to 6\% contribution to the total DOG $\rho_{\rm SFR} $. 

We note that \citet{Pope08} also examined  bump (star-forming) DOGs at $z\sim2$ down to $S_{24}=100\,\mu$Jy and they calculated $\rho_{\rm{SFR}}\sim1\times10^{-2} \text{ M}_{\odot}\text{yr}^{-1}\text{Mpc}^{-3}$, under the assumption that the average DOG has a star-formation rate of $200\,\rm{M}_{\odot}\rm{yr}^{-1}$. This value is lower than the bump $\rho_{\rm{SFR}} = 1.9\pm0.3 \times10^{-2}\text{ M}_{\odot}\text{yr}^{-1}\text{Mpc}^{-3}$ that we measure. However, it is difficult to determine whether these two values are significantly different because \citet{Pope08} do not provide an error on their measurement. We use their reported fractional error on the average IR luminosity (($1.1\pm0.7)\times10^{12}L_{\odot}$) to estimate that the minimum error on their $\rho_{\rm{SFR}}$ is $\sim0.6\times10^{-2}\text{ M}_{\odot}\text{yr}^{-1}\text{Mpc}^{-3}$, in addition to the contribution from the counting error from their 62 sources (compared to our 1137 bump sources at $z=1.5-2.5$). We also note that the selection criteria for the two studies are slightly different and if we were to use the bump DOG selection scheme in Pope et al.\,(2008; $S_{3.6} < S_{4.5} > S_{5.8}$ and $S_{4.5} > S_{8.0}$, or $S_{4.5} < S_{5.8} > S_{8.0}$ and $S_{3.6} >S_{8.0}$) we would identify 100 fewer bump DOGs (9\% of our sample of bump DOGs are at $z=1.5-2.5$). We conclude that the two results are consistent but since our study uses a larger sample and employs a combination of direct observations and redshift-binned stacking to determine our IR luminosities, we consider this measurement more accurate.

%%%%%%%%%%%%%%%%%%%%%%%%%
% Figure: SFR vs. Mass Relation
%%%%%%%%%%%%%%%%%%%%%%%%%
\begin{figure}
\begin{minipage}[b]{\linewidth}
\centering
\includegraphics[width=\textwidth]{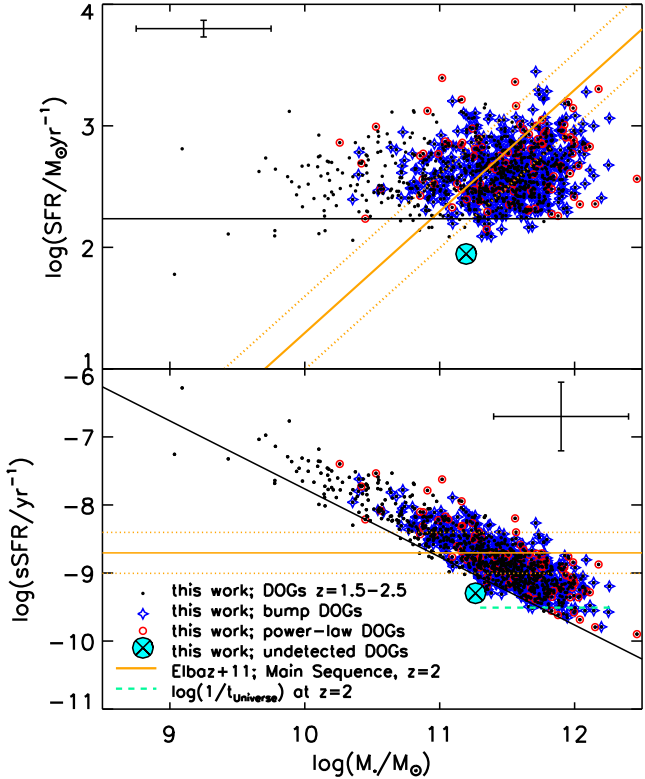}
\caption{Star-formation rate (SFR; {\it top panel}) and  specific star-formation rate (sSFR; {\it bottom panel}) as a function of stellar mass for DOGs at $z=1.5-2.5$. Power-law DOGs and bump DOGs are statistically indistinguishable in the $\text{SFR}-M_{\ast}$ plane. The sSFR at $z=2$ using the relation for star-forming galaxies from \citet{Elbaz11} and its conversion to SFR for the displayed range of masses is shown as the thick solid orange line in both panels. The orange dotted lines represent a factor of two dispersion from the derived SFR and sSFR. DOGs have a large scatter about the main sequence relation, having sources in the starburst, main sequence, and passive galaxy regimes. The thin horizontal black line in the top panel represents a minimum detectable star-formation rate at $z\sim2$, caused by the $24\,\mu$m flux density limit. Converting this to an sSFR value results in the diagonal line in the bottom panel, leading us to conclude that the apparent negative correlation between sSFR and stellar mass is a selection effect. }
\label{fig:sfrvsm}
\end{minipage}
\end{figure}
%%%%%%%%%%%%%%%%%%%%%%%%%
  
\subsection{Stellar Mass Build-up}
\label{ssec:smbu}
Using the stellar masses derived in \citet{Ilbert10} (corrected to assume a Salpeter IMF by adding +0.24 dex and be consistent with our SFR calculations), and our derived star-formation rates using {\it Herschel} data, we investigate where DOGs lie in the star-formation rate -- stellar mass ($\text{SFR} - M_{\ast}$) plane. Disk galaxies with a steady star-formation mode are observed to form a tight correlation in their star-formation rates as a function of stellar mass, defining a ``main sequence" \citep{Daddi07, Elbaz11}. Outliers above this relation are thought to be merger-driven starburst galaxies  (\citealt{Rodighiero11} and references therein). In the top panel of Fig.~\ref{fig:sfrvsm} we show the star-formation rates and stellar masses for {\it Herschel}-detected DOGs, considering only those at $z=1.5-2.5$ to minimize the effects of redshift evolution. Average error bars are plotted for star-formation rates and the uncertainties in stellar mass are fixed to 0.5 dex, which covers the systematic offset range due to the choice of extinction laws and stellar population synthesis models. 

Fig.~\ref{fig:sfrvsm} shows that power-law DOGs and bump DOGs cover the same ranges in stellar mass and star-formation rate in the $\text{SFR} - M_{\ast}$ plane, as expected if the far-IR is star-formation dominated. Our findings are also consistent with previous studies that investigated the similarities in properties of far-IR SEDs of {\it Herschel}-selected star-forming galaxies and AGN \citep{Mullaney12}.

The infrared main sequence from \citet{Elbaz11} for {\it Herschel}-selected star-forming galaxies at $z=2$ is also shown in Fig.~\ref{fig:sfrvsm}. DOGs have a significant amount of scatter about this relation, with $46\%$ within a factor of 2 of the main-sequence, $24\%$ above it and consistent with starbursts, and $31\%$ below it in the more quiescent regime. 

The bottom panel of Fig.~\ref{fig:sfrvsm} shows the specific star-formation rate ($\text{sSFR}=\frac{\text{SFR}}{M_{\ast}}$) as a function of stellar mass. The sSFR quantifies the weighted SFR and stellar mass, with its inverse giving the mass-doubling time for the current episode of star-formation activity. An apparent negative correlation in which lower mass $z=2$ DOGs exhibit higher sSFRs than their higher mass counterparts is observed, however this is largely a selection effect due to the flux limit of our sample. On the top panel of Fig.~\ref{fig:sfrvsm} we use the minimum IR luminosity at $z=2$ from our sample to represent a minimum detectable star-formation rate limit, shown as the horizontal line. We convert this to sSFR for a range of masses and this is shown as the diagonal line in the bottom panel. The logarithmic inverse age of the Universe in Gyr at $z=2$ is $\approx -9.5$ (dashed line in Fig.~\ref{fig:sfrvsm}) and most DOGs have sSFRs larger than this, indicating that the observed phase of star-formation could be responsible for their total observed stellar mass. 

Finally, we use the known redshift distribution and the sSFRs of DOGs to compare their volume densities to their proposed progenitors, SMGs. The volume density of observed DOGs with $S_{24}\ge100\,\mu$Jy at $z=1.5-2.5$ is $8\times10^{-5}\text{ Mpc}^{-3}$. Using the median DOG sSFR to estimate the characteristic lifetime of the DOG phase to be approximately 1 Gyr, we can correct this density for the burst duty cycle to derive a volume density for the progenitors to be approximately $10^{-4}\text{ Mpc}^{-3}$. This is consistent to the volume density for SMGs  at $z=1.5-3$ with $S_{870} > 4$ mJy derived from \citet{Wardlow11}, which assumes the lifetime of the SMG phase to be 100 Myr, 10 times shorter than for DOGs. In this scenario, DOGs would have the same descendants as $z\sim2$ SMGs, which are likely to be $2-3{ L^{*}}$ early-type galaxies \citep{Wardlow11,Hickox12}.  

\section{Conclusions}
We use {\it Herschel} HerMES data in COSMOS to study the far-IR emission from DOGs. The main findings are:
\begin{enumerate}
\item Out of 3077 DOGs, 51\% are detected in {\it Herschel} ($S_{250}\ge3\sigma$ = 8 mJy). We use stacking to probe the remaining {\it Herschel}-undetected population and the stacked $S_{250}$ is on average a factor of $\sim2$ fainter than the $250\,\mu$m detection limit of $3\sigma$. 

\item The IR luminosity functions of DOGs and all $24\,\mu$m sources with $S_{24} \ge 100\,\mu$Jy at $z=1.5-2.5$ are calculated. The stacked infrared luminosities provide significant contribution in the lowest {\it Herschel}-detected IR luminosity bin, causing an increase of $\sim0.2$ dex. IR luminosities derived from extrapolating $24\,\mu$m flux densities of CE01 templates are overestimated by a factor of 2 and in agreement with previous observations.

\item DOGs contribute $10-30\%$ to the overall star-formation rate density of the Universe and 30\% to all $24\,\mu$m galaxies with $S_{24}\ge100\,\mu$Jy. We also note that when compared to the total DOG $\rho_{\rm{SFR}}$, power-law (AGN dominated) DOGs provide minor contributions. The $\rho_{\rm{SFR}}$ for DOGs and SMGs are comparable at $z\sim2$, however we note that DOGs are more numerous, with individually lower star-formation rates for DOGs than SMGs. 

\item DOGs have a large scatter in the $\text{SFR}-M_{\ast}$ plane, having sources in the starburst, main sequence and more quiescent galaxy regimes. The observed phase of star-formation for most DOGs is likely responsible for their observed stellar mass. 

\end{enumerate}
\label{sec:conc}

We thank the anonymous referee for the useful comments that helped improve this paper. We thank Jason Melbourne for insightful discussions.

SPIRE has been developed by a consortium of institutes led by Cardiff Univ.\,(UK) and including: Univ. Lethbridge (Canada); NAOC (China); CEA, LAM (France); IFSI, Univ. Padua (Italy); IAC (Spain); Stockholm Observatory (Sweden); Imperial College London, RAL, UCL-MSSL, UKATC, Univ. Sussex (UK); and
Caltech, JPL, NHSC, Univ. Colorado (USA). This development has been supported by national funding agencies: CSA (Canada); NAOC (China); CEA, CNES, CNRS (France); ASI (Italy); MCINN (Spain); SNSB (Sweden); STFC, UKSA (UK); and NASA (USA). 

This research has made use of data from the HerMES project (\textcolor{blue}{\hyperlink{link:hermes}{http://hermes.sussex.ac.uk/}}). HerMES is a Herschel Key Programme utilizing Guaranteed Time from the SPIRE instrument team, ESAC scientists and a mission scientist. The data presented in this paper will be released through the HerMES Database in Marseille, HeDaM (\textcolor{blue}{\hyperlink{link:hedam}{http://hedam.oamp.fr/HerMES/}}).

We our thankful to the COSMOS collaboration for granting us access to their catalogs; and we gratefully acknowledge the contributions of the entire COSMOS team that have made
this work possible. More information on the COSMOS survey is available at \textcolor{blue}{\hyperlink{link:cosmos}{http://hermes.sussex.ac.uk/}}.
This work is based (in part) on observations made with the {\it Spitzer Space Telescope}, which is operated by the Jet Propulsion Laboratory, California Institute of Technology under a contract with NASA. 
We acknowledge support from a GAANN fellowship (to JAC), NSF CAREER AST-0645427 (AC an HF) and NASA funds to the US HerMES team through JPL.
S.O. acknowledges support from the Science and Technology Facilities Council [grant number ST/I000976/1].

Some of the data (spectroscopic redshifts) presented here were obtained at the W.M. Keck Observatory, which is operated as a scientific partnership among the California Institute of Technology, the University of California and the National Aeronautics and Space Administration. The Observatory was made possible by the generous financial support of the W.M. Keck Foundation. The authors wish to recognize and acknowledge the very significant cultural role and reverence that the summit of Mauna Kea has always had within the indigenous Hawaiian community. We are most fortunate to have the opportunity to conduct observations from this mountain.

{\it Facilities}: {\it Herschel} (SPIRE), {\it Spitzer} (IRAC, MIPS), Subaru (Suprime Cam), Keck
 
%\bibliographystyle{apj_hack}
%\bibliography{DOGS_paper.bib}

\end{document}